\documentstyle[seceq,epsf,preprint]{ptptex}
\title{%        %You can use \\ for explicit line-break
 4 D Quantum N-Dilaton Gravity\\ 
and \\
One-Loop Divergence of Effective Action \\
on Constant Dilaton
}
\author{%       %Use \sc for the family name
Hiroyuki {\sc Takata}
\footnote{E-MAIL: takata@theory.kek.jp, URL: http://theory.kek.jp/~takata/} 
}
\inst{%         %Affiliation, neglected when [addenda] or [errata]
National Laboratory for High Energy Physics (KEK),
Tsukuba, Ibaraki 305, Japan.\\
    and\\
Department of Physics, Hiroshima University,
Higashi-Hiroshima 739, Japan.
}
\recdate{%      %Editorial Office will fill in this.
\today
}
\abst{%         %this abstract is neglected when [addenda] or [errata]
We consider 4D quantum gravity with N-dilatons with the most general couplings.
Especially, on constant dilaton and arbitrary metric background,
we show the structure  of the divergent terms.
We show the constraint between the couplings
necessary to cancel the coefficient of the square of the Wyle tensor.
Next we show the N dependence of a  non-renormalizable divergent term,
and found that it cannot be canceled in the case of $N \geq 1$
with any fine-tuning of the couplings.
}
\begin{document}
\maketitle
\section{Introduction and Summary}
%\hspace*{5mm}
Recently there has been considerable interest in the metric-scalar  gravity in
four dimensions  from various point of view.
Our point of view is that such a fundamental theory of quantum gravity is
either a string theory or another theory like it.
We will not resolve this issue.
Furthermore, we do not remove the possibility
that  quantum gravity is  a local field theory with renormalizability.
%%%%%%%%%%%%%%%%%%%%%%%%%%%%%%%%%%%%%%%%%%%%%%%%%%%%%%%%%%%%%%%%%%%%%%%%
A candidate for a consistent theory of quantized gravity is string theory.
A low-energy effective theory of a string below the Plank scale represented
by a metric and a dilaton is well known \cite{GSW}.
Such an effective action arises in the form of a power-series type
of slope parameter ($\alpha^{\prime}$);
the standard point of view is
that the higher orders in such an expansion correspond to higher energies.
From this point of view, at a lower energy scale
the action for gravity has the form of a lower derivative dilaton action.
Since the fundamental theory is not restricted to the string theory,
we introduce N-dilations with the most general coupling to metric
within two derivatives, such as (\ref{action}).
Although in the string theory there is no scalar field having such a coupling,
we call our scalar fields dilatons by analogy.
The Einstein gravity coupled to scalars is nonrenormalizable
as naive power counting,
and higher derivative gravity is  renormalizable \cite{St};
however, it is not unitary within a perturbation scheme \cite{BOS}.
Of course, it is not strange
that a useful local field theory of gravity covering all energy regions
does not exist.
It is important to know whether a renormalizable local field theory of gravity
constructed by metric exists or not,
and what type of environment would allow its existence.
%%%%%%%%%%%%%%%%%%%%%%%%%%%%%%%%%%%%%%%%%%%%%%%%%%%%%%%%%%%%%%%%%%%%%%%%
%%%%%%%%%%%%%%%%%%%%%%%%%%%%%%%%%%%%%%%%%%%%%%%%%%%%%%%%%%%%%%%%%%%%%%%
Several studies, starting in seventies,
have been  performed to calculate the divergence of an effective action
of four-dimensional gravity \cite{HV,CD,BKK,ST1,Ta1,MT}.
In the pure Einstein action case without a cosmological term,
it was originally calculated at the one-loop level
by t'Hooft and Veltman \cite{HV}.
They found that the action is not renormalizable off mass shell,
but is finite on mass shell at the one-loop level.
Furthermore, although the pure Einstein action with a cosmological constant
is renormalizable \cite{CD},
if one introduces matter fields the one loop renormalizability is lost,
even on mass shell.
Recently \cite{ST1,ST2}
we considered the divergence of the effective action,
which is the most general class with less than two derivatives
for a scalar and a metric,
while explicitly leaving the functions $A,B,\Lambda$ arbitrary.
On an arbitrary back-ground space-time we found models
which are finite in the case without a cosmological term,
and with it are renormalizable by fine-tuning of functional form
of $A(\phi), B(\phi), \Lambda(\phi)$ at the one loop level on mass shell.
We have considered that on maximally symmetric background space-time
the action (\ref{action}) with $N=1$.
Without any fine-tuning of the coupling functions $A(\phi)$, $B(\phi)$,
we have shown that the divergence of the effective action has one term only
which proportional to $\Lambda^2$,
and the divergence can be renormalized easily.
%%%%%%%%%%%%%%%%%%%%%%%%%%%%%%%%%%%%%%%%%%%%%%%%%%%
In the present paper we consider the action:
%%%%%%%%%%%%%%%%%%%%%%%%%%%%%%
\begin{equation}
S\left[g_{\mu\nu},\phi_i \right]
= \int d^4x \sqrt{-g}\;
\left[ A(\phi)_{i j}g^{\mu\nu}\partial_{\mu}\phi_i
\partial_{\nu}\phi_j + B(\phi)R -2B(\phi) \Lambda(\phi) \right]
\;\;\;i=1 \cdots N
\label{action}
\end{equation}
%%%%%%%%%%%%%%%%%%%%%%%%%%%%%%
This is the most general class with less than two derivative for N scalars
and a metric.
Since by redefinition of fields $A_{i j} \longrightarrow A \delta_{i j}$
in generic,
in this paper we consider $A_{i j}=A \delta_{i j}$ case only.
%%%%%%%%%%%%%%%%%%%%%%%%%%%%%%
\begin{equation}
S\left[g_{\mu\nu},\phi_i \right]
= \int d^4x \sqrt{-g}\; \left[ A(\phi)g^{\mu\nu}\partial_{\mu}\phi_i
\partial_{\nu}\phi_i + B(\phi)R -2B(\phi) \Lambda(\phi) \right]
\;\;\;i=1 \cdots N
\label{action2}
\footnote{
In this paper we restrict
$B \neq 0$ and $A \neq \frac{3}{2}\frac{B_i B_i}{B}$,
where we write $X_{i_1 \cdots i_n}
:=\frac{ \partial ^n X(\phi)}{\partial \phi_{i_1} \cdots \partial \phi_{i_n}}$,
$X_i X_i := \sum_i^{N} X_i X_i$ for any function $X(\phi)$}
\end{equation}
%%%%%%%%%%%%%%%%%%%%%%%%%%%%%%%%%%%%%%%%%%%%%%%%%%%%%%%%%%%%%%%%%%%%%%%
Our paper is organized as follows.
In section 2,
we consider the classical analysis of the action (\ref{action}).
We show classically the non-equivalence
between the class of  action (\ref{action})
and the class of the action without the kinetic term of the dilatons
in (\ref{action}).
In section 3,
we calculate the divergence of the effective action
with the background field method and the Schwinger-Dewitt method.
Especially, on constant dilaton background, we show an explicit calculation,
and we get the structure of the divergence.
In section 4,
we restrict the form of the couplings
in order to cancel a non-renormalizable term,
and we show $N$ dependence of another non-renormalizable term
which cannot be canceled in the case of $N \geq 1$.
In section 5,  we conclude this paper.
We have three Appendixes.
\section{Analysis at the Classical Level}
%\hspace*{5mm}
We consider gravity with a general coupling to scalars
in which the action is (\ref{action}).
In this section we analyze this theory at the classical level.
\subsection{Classical Non-Equivalence
between Constant and Non-Constant Dilaton Cases }
%\hspace*{5mm}
In a previous paper\cite{ST1} which treated the  $N=1$ case in (\ref{action}),
we have shown the equivalence
between an original action and the no kinetic term action.
In this subsection, however, we show for $ N > 1$ a classical non-equivalence
between the original action (\ref{action2})
and a model without kinetic term of dilaton ($\partial \phi_i = 0$)
in the original action (\ref{action2}).
First we start with an action without kinetic terms of dilatons:
\begin{equation}
S\left[ \bar{g}_{\mu\nu},\phi_i \right]=
\int d^4x \sqrt{-\bar{g}}
\; \left[ {\cal B}(\phi)\bar{R} -2  {\cal B}(\phi)  \lambda(\phi) \right]\;\;\;
\label{stand}
\end{equation}
We transform the metric:
\begin{equation}
\bar{g_{\mu\nu}} \longrightarrow g_{\mu\nu}=e^{2 \sigma(\phi)}\bar{g_{\mu\nu}}
\end{equation}
where $\sigma(\phi)$, ${\cal B}(\phi)$ and  $\lambda(\phi)$
are arbitrary functions of $\phi$.
In a new variable the action becomes:
\[
S\left[ g_{\mu\nu},\phi_i \right]
= \int d^4x \sqrt{-g} \times
\]
{\small
\begin{equation}
\left[
6 e^{2\sigma(\phi)}
\left({\cal B} \sigma_i \sigma_j
 + \frac{1}{2}\left({\cal B}_i \sigma_j
 +\sigma_i{\cal B}_j \right) \right)(\nabla \phi_i)(\nabla \phi_j)
+ {\cal B} (\phi) e^{2\sigma(\phi)} R
-2 {\cal B}(\phi) \lambda(\phi) e^{4\sigma(\phi)} \right]
\footnote{We use the convinient notations:
$(\nabla \phi_i)(\nabla \phi_j)
= g^{\mu \nu}(\partial_\mu \phi_i)(\partial_\nu \phi_j)$
and
$(\nabla \phi)^2 =(\nabla \phi_i)(\nabla \phi_i)$  }
\end{equation}
}
If we can set $\sigma(\phi)$, ${\cal B}(\phi)$ and $ \lambda(\phi)$ to
\begin{equation}
6 e^{2\sigma(\phi)}\left({\cal B} \sigma_i \sigma_j
+ \frac{1}{2}\left({\cal B}_i \sigma_j +\sigma_i{\cal B}_j \right) \right)
=A(\phi)_{i j}
\;,\;\;\;
{\cal B}(\phi)e^{2\sigma(\phi)}= B(\phi)
\;,\;\;\;
\frac{ \lambda(\phi)}{{\cal B}(\phi)}=\frac{\Lambda(\phi)}{B(\phi)}
\label{functs}
\end{equation}
for arbitrary functions $A(\phi)_{i j}$, $B(\phi)$ and $\Lambda(\phi)$,
then the original action (\ref{action}) and the action (\ref{stand})
are  equivalent.
If $N > 1 $ and $A_{i j}$ is diagonal, however,
the first equation in (\ref{functs}) cannot be satisfied
except for $A_{i j}=0$.
This is an essential difference from the $N=1$ case.
Therefore we will analyze the model in the case of $N > 1$.
\begin{equation}
\end{equation}
\subsection{Classical Equations of Motion}
The classical equations of motion for $g_{\mu\nu}$ and $\phi_i$ are
%%%%%%%%%%%%%%%%%%%%%%%%%%%%%%
\begin{equation}
R_{\mu\nu}
-\frac{1}{2}R g_{\mu\nu}
+ \Lambda g_{\mu\nu}
=T_{\mu\nu} \;\;\;\;\;\;(\;\mbox{for}\;g_{\mu\nu})
\end{equation} 
and
\begin{equation}
 B_i R -2(B\Lambda)_i +A_i(\nabla \phi)^2 
-2 A_j(\nabla_{\mu} \phi_j)(\nabla_{\nu} \phi_i)
-2A (\Box \phi_i) 
=0 \;\;\;\;\;\;(\;\mbox{for}\;\phi_i)\;,
\end{equation}
where 
\[
T_{\mu\nu}:=
\]
\[
\left(
\frac{A}{2 B}(\nabla \phi)^2 
- \frac{B_{ij}}{B}(\nabla \phi_i)(\nabla \phi_j)
- \frac{B_i}{B}(\Box \phi_i) 
\right) g_{\mu\nu}
\]
\begin{equation}
+ \frac{B_{ij}}{B}(\nabla_{\mu} \phi_i)(\nabla_{\nu} \phi_j)
-\frac{A}{B} (\nabla_{\mu} \phi_i)(\nabla_{\nu} \phi_i)
+\frac{B_i}{B}(\nabla_{\mu}\nabla_{\nu}\phi_i)\;.
\end{equation}
%%%%%%%%%%%%%%%%%%%%%%%%%%%%%%
%%%%%%%%%%%%%%%%%%%%%%%%%%%%%%
Especially, we consider special solution  with the constant dilaton.
In that case, the energy momentum tensor vanishes and  the classical action is
\begin{equation}
S_{\partial \phi_=0}
= \int d^4x \sqrt{-g}\; \left[ B(\phi)R -2B(\phi) \Lambda(\phi) \right]\;.
\end{equation}
This is same to (\ref{stand})which is the action
with no kinetic term of dilaton.
The equations of motion are
\begin{equation}
R_{\mu\nu}
-\frac{1}{2}R g_{\mu\nu}
+ \Lambda g_{\mu\nu}
=0
\label{veomg}
\end{equation}
\begin{equation}
B_i R -2(B \Lambda)_i =0
\label{veomp}
\end{equation}
In Appendix A we consider solution classically.
%%%%%%%%%%%%%%%%%%%%%%%%%%%%%%
%\hspace*{5mm}
%%%%%%%%%%%%%%%%%%%%%%%%%%%%%%
\section{One-loop calculations}
\subsection{BackGround Field Method}
We consider the one-loop divergence of the effective action.
First, we start with the background field method \cite{Ab}. 
We split the fields
into background fields ($g_{\mu\nu}$, $\phi_i$)
and quantum fields ($h_{\mu\nu}$, $\varphi_i$):
\begin{equation}
\phi_i \rightarrow \phi_i^{\prime} 
= \phi_i + \varphi_i \;,\;\;\;\;\;\;\;\;\;\;\;\;\;\;
\;   
g_{\mu\nu} \rightarrow g'_{\mu\nu} = g_{\mu\nu} + h_{\mu\nu}
\end{equation}
Although the original action (\ref{action2}) and the action (\ref{stand})
are not equivalent when $N >1$ for the reason shown in the previous section,
when background dilatons are constant
the two classical actions have the same form.
Classically the theory of gravity is explained well by Einstein  action.
Therefore we set the classical background dilaton $\phi_i$ to be constant
while the quantum fluctuation $\varphi_i$ is allowed to vary.
On the other hand we do not restrict the background and quantum metric. 
Since the action (\ref{action2}) has diffeomorphic invariance
we have to fix the gauge freedom.
We fix the quantum field with the gauge fixing term:
\begin{equation}
S_{gf} = \int d^4 x \sqrt{-g}\;\chi_{\mu}\;\frac{\alpha}{2}\;\chi^{\mu}          
\end{equation}
where
\footnote{$h=h_{\mu}^{\mu},\;
\bar{h}_{\mu\nu}
=h_{\mu\nu}-\frac{1}{4}\;hg_{\mu\nu}$}
\begin{equation}
\chi_{\mu}
= \nabla_{\alpha} \bar{h}_{\mu}^{\,\alpha}+
\beta\nabla_{\mu}h 
+ \gamma_{i} \nabla_{\mu} \varphi_{i}\;.
\end{equation}
are functions of the background dilaton.\\
In order to simplify the differential structure of the bilinear part
of the total action ($S+S_{\mbox{gf}}+ S_{\mbox{gh}}$ ),
we choose these functions
as
\begin{equation}
\alpha=-B\;\;,\;\;\;\;\beta=-\frac{1}{4}\;\;,\;\;\;\;\gamma_i=-\frac{B_i}{B}\;,
\end{equation}
which induces
\begin{equation}
\left.\left(S + S_{\mbox{gf}} +S_{\mbox{gh}}\right)\right|_{\mbox{bilinear}}
=\int d^4 x \sqrt{-g}\;
\left({\Phi} \hat{H} {\Phi}^T + c_{\mu}\hat{H}_{\mbox{gh}c^{\mu}}
\right)\;,
\end{equation}
where
\[
 \hat{H}
=\hat{K}\Box
+\hat{L}_{\rho}\nabla^{\rho}
+ \hat{M}\;,
\]
\begin{equation}
\hat{H}_{\mbox{gh}}
= g^{\mu\alpha}\Box
+\gamma_i(\nabla^{\alpha}\phi_i)\nabla^{\mu}
+ \gamma_i(\nabla^{\mu} \nabla^{\alpha} \phi_i)
+ R^{\mu \alpha}
\end{equation}
%%%%%%%%%%%%%%%%%%%%%%%%%%%%%%
Here, $\Phi=\left(\bar{h}_{\mu\nu},\;h,\; \varphi\right)$ and  $c_{\mu}$
stand for  ghosts and $T$ stands for transposition.\\
The components of $\hat{H}$ have the following form:
%%%%%%%%%%%%%%%%%%%%%%%%%%%%%%
\begin{equation}
\hat{K}=\left(
\begin{array}{ccc}
              \frac{B}{4} \delta^{\mu\nu \alpha \beta} & 0 & 0\\
              0 & -\frac{B}{16} & -\frac{B_j}{4} \\
              0 & -\frac{B_i}{4} & \frac{B_iB_j}{2B} -A \delta_{ij}
\end{array}
\right)
\footnote{$\delta^{\mu\nu \alpha \beta}
:= \frac{1}{2}\left( g^{\mu \alpha}g^{\nu \beta}
+g^{\mu \beta}g^{\nu \alpha}\right)$ }
\end{equation}
\[
\hat{L}^{\lambda}=(\nabla_{\tau}\phi_k) \times
\]
{\scriptsize
\begin{equation}
\left(\!\!\!\!\!\!\!
\begin{array}{ccc}
              \frac{B_k}{4} \left(\delta^{\mu \nu \alpha \beta}
g^{\tau \lambda}
+2 g^{\nu \beta}\left(g^{\mu \tau } g^{\alpha \lambda }
             - g^{\alpha \tau } g^{\mu \lambda }\right) \!\! \right)

\!\!\!\!\!\!\!
&
\!\!\!\!\!\!\!

 - \frac{B_k}{4} g^{\mu \tau} g^{\nu \lambda}
       
\!\!\!\!\!
&
\!\!\!\!\!

 \left( \frac{B_{jk}}{2}-A\delta_{jk} \right) g^{\mu \tau} g^{\nu \lambda}\\
           \frac{B_k}{4} g^{\alpha \tau} g^{\beta \lambda}
            
\!\!\!\!\!\!\!
&
\!\!\!\!\!\!\!

 -\frac{B_k}{16} g^{\tau \lambda}
        
\!\!\!\!\!
&
\!\!\!\!\!

 \left(\frac{A}{4}\delta_{jk} -\frac{5}{8} B_{jk} \right) g^{\tau \lambda}\\
   \left( A\delta_{ik} - \frac{B_{ik}}{2}\right) g^{\alpha \tau} g^{\beta \lambda}
            
\!\!\!\!\!\!\!\!
&
\!\!\!\!\!\!\!\!

 \left( \frac{B_{ik}}{8}-\frac{A}{4}\delta_{ik}\right) g^{\tau \lambda}
           
\!\!
&
\!\!

 \left( \!\!\! \left( \!\! \frac{B_i B_j}{2B} - A\delta_{ij} \!\! \right)_{\!\!k} \!\!\! \left(  A_i\delta_{jk}-A_j\delta_{ik}\right) \!\!\! \right) g^{\tau \lambda}
\end{array}
\!\!\!\!\!\!\! \right) 
\end{equation}
}
\[
\hat{M}=
\]
{\scriptsize
\begin{equation}
 \left(\!\!\!\!\!\!\!\!\!
\begin{array}{ccc}
    \begin{array}{l}
                   \delta^{\mu \nu \alpha \beta}\left( \frac{B_k}{2}
                                                (\Box \phi_k)  
                 + \left( \frac{B_{kl}}{2}-\frac{A}{4}\delta_{kl} \right) 
                                      (\nabla \phi_k)^(\nabla \phi_l)
                   + \frac{B \Lambda}{2} \right)
               \\  + g^{\nu \beta}\left(
          -B_k\left( \nabla^{\mu} \nabla^\alpha \phi_k  \right)
   +\left( A\delta_{kl}-B_{kl} \right)(\nabla^\mu \phi_k)(\nabla^\alpha \phi_l)\right)
           \\ +\frac{B}{4}\left( -\delta^{\mu \nu \alpha \beta} R
 + 2 g^{\nu \beta} R^{\mu \alpha}+2R^{\mu \alpha \nu \beta} \right)
           \end{array}
    
           \!\!\!\!\!\!\!\!\!\!\!\!\!\!\!\!
          & \!\!\!\!\!\!\!\!\!\!\!\!\!\!\!\! 0 \!\!\!\!\!\!\!\!\!\!\!\!\!\! 
          & \!\!\!\!\!\!\!\!\!\!\!\!\!\!

   \begin{array}{l}
          \frac{B_{jk}}{2}\left( \nabla^{\mu} \nabla^{\nu} \phi_k \right)
              \\
               + \left( \frac{B_jkl}{2} - \frac{A_j}{2}\delta_{kl} \right)
                                  (\nabla^\mu \phi_k)(\nabla^\nu \phi_l)
              \\
                - \frac{B_j}{2}R^{\mu \nu}
    \end{array}

 \\ \!\!\!\!\!\!\!\! & \!\!\!\!\!\!\!\!  \!\!\!\!\!\! & \!\!\!\!\!\!\!\!
          
 \\
            \frac{B_k}{4}\left( \nabla^{\alpha} \nabla^{\beta} \phi_k \right)
              + \frac{B_{kl}}{4} (\nabla^\alpha \phi_k)(\nabla^\beta \phi_l)

           \!\!\!\!\!\!\!\!\!\!\!\!\!\! & \!\!\!\!\!\!\!\!\!\!\!\!\!\!
            -\frac{B \Lambda}{8} 
           \!\!\!\!\!\!\!\! & \!\!\!\!\!\!\!\!\!\!\!\!
            \begin{array}{l}

               -\frac{3}{8} B_{jk} (\Box \phi_k)
     
               \\
              + \left( \frac{A_j}{8}\delta_{kl} 
              - \frac{3}{8}B_{jkl} \right)(\nabla \phi_k)(\nabla \phi_l)
                \\
             
              + \frac{B_j}{8} R - \frac{(B \Lambda)_j}{2}
             
     \end{array}
 \\ 
 \!\!\!\!\!\!\!\! & \!\!\!\!\!\!\!\!\!\!\!\!\!\! &
 \!\!\!\!\!\!\!\! 
 \\
             A\left( \nabla^{\alpha} \nabla^{\beta} \phi_i \right)
             + \frac{A_i}{2} (\nabla^\alpha \phi)(\nabla^\beta \phi)
              - \frac{B_i}{2}R^{\alpha \beta}

            \!\!\!\! & \!\!\!\!
    
     \begin{array}{l}
     
             -\frac{A}{4} (\Box \phi_i)
    
                \\
        
            +\frac{A_i}{8} (\nabla \phi)^2
             
                 \\

            -\frac{A_k}{4}(\nabla \phi_k)(\nabla \phi_i)

                \\
               
            + \frac{B_i}{8} R - \frac{(B \Lambda)_i}{2}
           
     \end{array}
  
              \!\!\!\!\!\!\!\! & \!\!\!\!\!\!\!\!\!\!
    
      \begin{array}{l}
    
             -A_j(\Box \phi_i)
   
                  \\
       
            +\frac{A_{ij}}{2}(\nabla \phi)^2
      
                   \\
          
              -A_{jk}(\nabla \phi_k)(\nabla \phi_i)
             
                   \\

             + \frac{B_{ij}}{2} R - (B \Lambda)_{ij} 
         
      \end{array}
\end{array}
\!\!\!\!\!\!\!\!\!\! \right)
\end{equation}
}
The one loop effective action is given by the standard general expression,
%%%%%%%%%%%%%%%%%%%%%%%%%%%%%%
\begin{equation}
\Gamma^{\mbox{\small 1-loop}}={i \over 2}\;\mbox{Tr} \ln {\hat{H}}
- i\;\mbox{Tr}\ln {\hat{H}_{\mbox{gh}}}
\footnote{ Tr includes space time integral, tr does not. },
\end{equation}
%%%%%%%%%%%%%%%%%%%%%%%%%%%%%%%
\subsection{Schwinger-DeWitt Formula}
In this subsection we use the version of the the  Schwinger-DeWitt formula
for the case of constant N-dilaton.
For our minimal gauge, there are no second derivative term
except for a d'Alembertian term, for which is the convenient formula
of the structure of the divergence of the one loop effective action.
In Appendix B we present a  short review
of the Schwinger-DeWitt formula\cite{De,BV,BOS}.
We apply  this formula to our case.
From now on  we restrict the background dilatons to be constant
in order to simplify the calculation.
Note that the quantum fluctuation $(\varphi )$ of $\phi$
is not restricted to a constant.
There are no restriction on the metrics ($g_{\mu \nu}$ and $h_{\mu \nu}$).\\
After some calculations,
\begin{equation}  
\hat{P}=
\left(
\begin{array}{cc}
 D_{\mu \nu \alpha \beta} 
+ \left( \frac{R}{6} + 2 \Lambda \right) \delta_{\mu \nu \alpha \beta}
&
p_{1 2}R_{\mu \nu} 
\\
p_{2 1}R_{\alpha \beta}
& 
p_r R + p_l
\end{array}
\right)
\end{equation}
where
\begin{equation}
\left\{
\begin{array}{ll}
D_{\mu \nu \alpha \beta}=  
2 R_{\mu \alpha \nu \beta}
+ 2 g_{\nu \beta} R_{\mu \alpha}
- R \delta _{\mu \nu \alpha \beta}
&     
\\
p_{1 2}= \left(\frac{4}{B} \;\;\; - \frac{2 B_i}{B} \right)\;\;,
&
p_{2 1}= k^{-1}\left(\begin{array}{c} 0 \\ -\frac{B_i}{2}\end{array}\right)  
\\
p_r = 
k^{-1}\left(\begin{array}{cc} 0 & \frac{B_j}{8} 
                                     \\
                                      \frac{B_i}{8} & \frac{B_{i j}}{2}
            \end{array} \right)                  
+      \left(\begin{array}{cc} \frac{1}{6} & 0 
                                     \\
                               0 & \frac{1}{6}
            \end{array} \right)\;\;,
&
p_l=
k^{-1}\left(\begin{array}{cc} -\frac{B \Lambda}{8} & -\frac{(B \Lambda )_j}{2} 
                                     \\
                               -\frac{(B \Lambda )_i}{2}  & -(B \Lambda)_{i j}
            \end{array} \right)
\\
k= 
\left(\begin{array}{cc} -\frac{B}{16} & -\frac{B_j}{4} 
                                     \\
                        -\frac{B_i}{4}  & \frac{B_i B_j}{2 B}-A \delta_{i j}
            \end{array} \right)
& 
\end{array}
\right.\footnote{$k^{-1}$ exists when $X:=2 A B -3 B_i B_i \neq 0 $}
\end{equation}
\begin{equation}
\hat{S}_{\lambda \lambda^{\prime}}=
\left(
\begin{array}{cc}
2 g_{\nu \beta}R_{\mu \alpha \lambda \lambda^{\prime}}
&
0
\\
0
& 
0
\end{array}
\right)
\end{equation}
\begin{equation}
\hat{P}_{\mbox{gh}}=R_{\mu \alpha} + \frac{R}{6} g_{\mu \alpha}
\end{equation}
\begin{equation}
\hat{S}_{\mbox{gh};\lambda \lambda^{\prime}}=R_{\alpha \mu \lambda \lambda^{\prime}}\;.
\end{equation}
Therefore, the divergence of the one-loop effective action
with constant background dilatons is
\[
\Gamma_{\mbox{div },\; \partial \phi=0}^{\mbox{1-loop}}
= \frac{1}{16 \pi^2 (D-4)} \int d^4 x \sqrt{-g} \times 
\]
\[
\left[
\frac{N+212}{180}R_{\mu \nu \alpha \beta}R^{\mu \nu \alpha \beta} 
+\left( p_{1 2} p_{2 1} - \frac{N+722}{180}\right) R_{\mu \nu}R^{\mu \nu}
\right.
\]
\begin{equation}
\left.
+\left( \frac{1}{2} \mbox{tr}p_r^2 
          -\frac{1}{4} p_{1 2} p_{2 1} 
          + \frac{85}{72} \right) R^2 
+\left( \frac{9}{2}\Lambda + \mbox{tr}p_r p_l \right) R
+\left( \frac{9}{2}\lambda^2 + \frac{1}{2} \mbox{tr}p_l^2 \right)
\right]\;.
\end{equation}
For convenience, we write the above expression
with the Weyl tensor ($C_{\mu \nu \alpha \beta}$)
and Gauss-Bonnet topological invariant quantity ($ G $):
\footnote{$G \equiv
R_{\mu\nu\alpha\beta}R^{\mu\nu\alpha\beta}-4R_{\mu\nu}R^{\mu\nu}+R^2$,
$\;C_{\mu\nu\alpha\beta}C^{\mu\nu\alpha\beta} \equiv
R_{\mu\nu\alpha\beta}R^{\mu\nu\alpha\beta}-2R_{\mu\nu}R^{\mu\nu}
+\frac{1}{3}R^2 $ }.
\[
\Gamma_{\mbox{div},\; \partial \phi=0}^{\mbox{1-loop}}
= \frac{1}{16 \pi^2 (D-4)} \int d^4 x \sqrt{-g} \times 
\]
\[
\left[
\left( \frac{1}{2} p_{1 2} p_{21} + \frac{298 -N}{360}\right)G 
+\left( \frac{1}{2 } p_{1 2} p_{2 1} + \frac{N+42}{120}\right)C_{\mu \nu \alpha \beta}C^{\mu \nu \alpha \beta}
\right.
\]
\begin{equation}
\left.
+\left( \frac{1}{2} \mbox{tr}p_r^2 
          +\frac{1}{12} p_{1 2} p_{2 1} 
          + \frac{17}{72} \right) R^2 
+\left( \frac{9}{2}\Lambda + \mbox{tr}p_r p_l \right) R
+\left( \frac{9}{2}\Lambda^2 + \frac{1}{2} \mbox{tr}p_l^2 \right)
\right]\;.
\label{effact}
\end{equation}
%%%%%%%%%%%%%%%%%%%%%%%%%%%%%%
\section{Removing the Non-Renormalizable Divergent Terms}
%\hspace*{5mm}
We consider the divergent term in the equation (\ref{effact}).
In (\ref{effact}) two divergent terms,
the scalar curvature term and cosmological term, appear in the classical action,
therefore its counter-terms are arranged.
However, in (\ref{effact}), there are first three terms
which cannot be canceled by the counter-terms.
First, we fine tune the functions $A(\phi)$
in order to cancel the coefficient of the quadratic term in the Wyle tensor.
Since $p_{1 2}p_{2 1}$ is calculated as
%%%%%%%%%%%%%%%%%%%%%%%%%%%%%% 
\begin{equation}
p_{1 2}p_{2 1}= -\frac{2 B_i B_i}{2A B -3 B_i B_i}
\end{equation}
%%%%%%%%%%%%%%%%%%%%%%%%%%%%%%
we set $A(\phi)$ to
%%%%%%%%%%%%%%%%%%%%%%%%%%%%%%
\begin{equation}
A(\phi)=\frac{3}{2}\left( 1 + \frac{40}{N + 42}\right)\frac{B_i B_i}{B}\;\;.
\end{equation}
%%%%%%%%%%%%%%%%%%%%%%%%%%%%%%
Remark: When $N$ tends to infinity,  $A$ is of the same form
as the  conformal symmetric case\cite{ST2}.
Since the coefficient of the Gauss-Bonnet term is constant in this case,
this term is total derivative.
The divergence of the surface term  is non-essential and is ignored.
Last problem is to consider
the divergence of the square of the scalar curvature term.
After the fine tuning of the function $A(\phi)$, this term is reduced to
%%%%%%%%%%%%%%%%%%%%%%%%%%%%%%
\[
\left[ \frac{1}{2} \mbox{tr}p_r^2 
          +\frac{1}{12} p_{1 2} p_{2 1} 
          + \frac{17}{72} \right] R^2 
\]
\[
=\left[\frac{N^2 +224 N +5344}{3600}
-\frac{(N-1)(N+42)}{18(N+82)}\left(\frac{B}{\phi B^{\prime}} \right) 
+\frac{(N-1)(N+42)^2}{18(N+82)^2}\left(\frac{B}{\phi B^{\prime}} \right)^2
\right.
\]
\begin{equation}
\left.-\frac{(N+42)(N+52)}{7200}\left( \frac{B B^{\prime \prime}}{B^{\prime 2}}\right)
+\frac{(N+42)^2}{28800}\left( \frac{B B^{\prime \prime}}{B^{\prime 2}}\right)^2
\right]R^2
\label{coefrr}
\footnote{$\phi:=\left(\phi_i \phi_i \right)^{\frac{1}{2}}$
and $B^{\prime}= \frac{\phi_i B_i}{\phi}\;,
\;B^{\prime \prime}= B_{ii}+(1-N)\frac{\phi_i B_i}{\phi^2}$ the primes mean differentiations respect to $\phi$ if $B(\phi)$ is the function of only $\phi$}\;\;.
\end{equation} 
%%%%%%%%%%%%%%%%%%%%%%%%%%%%%%
Fig.\ref{fig} in Appendix C we show the parameter region
where the coefficient of (\ref{coefrr}) vanishes.
We found that when $N \geq 1$ the coefficient cannot vanish
and this model is non-renormalizable at one loop in our method.
%%%%%%%%%%%%%%%%%%%%%%%%%%%%%%
%%%%%%%%%%%%%%%%%%%%%%%%%%%%%%
%\vspace*{\fill}
\section{Conclusion and Discussion}
%\hspace*{5mm}
We considered the model which includes N-scalar fields and a metric field.
First we analyze this model at the classical level.
At the classical level and in the case of only $N=1$,
the action (\ref{action2}) reduces to some standard form
by a conformal transformation.
However, in the case of $N>1$, there are no such equivalence.
Therefore introduction of the dilatons has essential meaning in $N>1$  case.
On the other hand,
the standard form (\ref{stand}) belongs to the class
without the kinetic term of dilatons in the original action (\ref{action2}).
There is also no equivalence at the quantum level between such models.
We restrict, however, back ground classical field to the constant dilaton since
the Einstein gravity explains well the nature at the classical level,
while the quantum fluctuations of dilatons is allowed to vary.
Of course the classical and quantum metrics do not have any restrictions.
%%%%%%%%%%%%%%%%%%%%%%%%%%%%%%
A one-loop calculation was carried out for the model (\ref{action2})
using the background field method.
This calculation is an extension of the case of Ref.\cite{ST1}.
%%%%%%%%%%%%%%%%%%%%%%%%%%%%%
We pulled a bilinear form out of the action (\ref{action2})
with a gauge fixing term added, and out of the ghost action.
Such a form is sufficient to calculate the effective action at one-loop level.
We have fixed a gauge to the minimal one
in order to cancel the derivative terms,
except for the d'Alembertian terms,
we were then able to apply the standard Schwinger-DeWitt method
to estimate the divergence of the effective action.
%%%%%%%%%%%%%%%%%%%%%%%%%%%%
We got a one-loop divergent term (\ref{effact}).
There are naively three non renormalizable terms.
However when we fine tune the function $A(\phi)$
there is only one non-renormalizable term which is the $R^2$ term.
We  show  $N$ and $B(\phi)$ dependences explicitly.
And graphically we show the region where the term vanishes.
We found that there is no region when $N \geq 1$.
Therefore it is impossible to renormalize
the divergence of the effective action at the one-loop level
on a constant dilaton background.
If we consider the metric to be on mass shell,
divergent terms may be renormalized,
as shown in a previous paper\cite{Ta1} and an incoming paper\cite{MT},
which treats the $N=1$ case.
In the case of constant dilaton and $R_{\mu \nu}= \Lambda g_{\mu\nu}$,
we think that the last three terms in (\ref{coefrr})
are proportional to $\Lambda^2$ with constant of $\phi$,
as in the above Refs.
Then, by multiplicative renormalization of the function form of the $\Lambda$,
the divergences are renormalized:
\begin{equation}
\Lambda_{\mbox{bare}}
=\mu^{\frac{D-4}{2}}
\left(1 -\frac{\mbox{constant}}{16 \pi^2 (D-4)} \right)
\Lambda_{\mbox{renormalized}}\;\;.
\end{equation}
In this paper we considered the $A_{ij}=A \delta_{ij}$ case.
In this case we cannot arrange counter term if $N \geq 1$.
However, in our next studies, we have to consider more general case
such as that there is no redefinition of fields to allow us to set
$A_{ij} = A \delta_{ij}$.
One  cannot neglect the possibility of a renormalizable model
in the class that metric coupled to N-scalars in the most general way.
If such a general case, there may be also some models
which differ essentially from the standard form (\ref{stand}).
Such a  model also may not be equivalent to the standard form (\ref{stand}).
\vspace*{\fill}
\section*{Acknowledgements}
%\hspace*{5mm}
The author is grateful to H.Kawai and the entire Department of Theoretical
and Computational Physics at KEK for the stimulating discussions.
He is also grateful to I.L.Shapiro for various suggestions by e-mail.
He is thankful to T.Muta, S.Mukaigawa
and the entire Department of Particle Physics at Hiroshima University
for the interesting discussions.
\appendix
\section{A classical solution on constant dilatons}
There is a solution for the  equations of motion (\ref{veomg}) and (\ref{veomp}) 
when $B \propto \Lambda$.
The solution must satisfy
\begin{equation}
R_{\mu \nu}= \Lambda g_{\mu\nu}\;\;.
\end{equation}
This includes the maximally symmetric solution
and spherically symmetric black hole solution.
That is
\begin{equation}
g_{\mu \nu}
=
\left(
\begin{array}{cccc}
 - e^{2 \nu} & & & \\
 & e^{-2 \nu} & &  \\
 & & r^2 & \\
& & & r^2\sin{\theta}          
\end{array}
\right)
\left.
\begin{array}{cccc}
 \cdots t \\
\cdots r \\
\cdots \theta \\
\cdots \varphi
\end{array}
\right. ,
\end{equation}
where
\begin{equation}
e^{2 \nu(r)}
= \mbox{( $\pm 1$ or $0 $)} - \frac{G}{r} + \frac{E}{r^2} -\frac{\Lambda}{3} r^2
\;\;\;\;\mbox{(G,E are some constant)}
\end{equation}
When $G=E=0$, this solution  is the maximally symmetric case.
When $\Lambda=0$, it is the Reissner-Nordstrom black hole  case,
when $\Lambda=E=0$, it is the Schwarzschild  black hole  case.
The divergence of the one-loop effective action on mass shell
on background with  these solutions is miraculously cancelled
in the model\cite{Ta1,MT}.
%%%%%%%%%%%%%%%%%%%%%%%%%%%%%%
\section{Schwinger-DeWitt  formula}
We start by choosing the essentially divergent part
from $\mbox{Tr}\ln \hat{H}$ and $\mbox{Tr}\ln \hat{H}_{\mbox{gh}}$:
%%%%%%%%%%%%%%%%%%%%%%%%%%%%%%
\begin{equation}
\mbox{Tr} \ln\hat{H} =\mbox{Tr}\ln\hat{K}+ \mbox{Tr}\ln\left(\hat{1}\Box +
\hat{K}^{-1} \hat{L}^{\mu}\nabla_\mu +\hat{K}^{-1}\hat{M} \right)\;.
\label{trlnH}
\end{equation}
%%%%%%%%%%%%%%%%%%%%%%%%%%%%%%
We define $\hat{E}$, $\hat{D}$, $\hat{E}_{\mbox{gh}}$ and $\hat{D}_{\mbox{gh}}$:
%%%%%%%%%%%%%%%%%%%%%%%%%%%%%%
\begin{equation}
\hat{E}^{\mu}:=\hat{K}^{-1} \hat{L}^{\mu}\;\;,\;\;\hat{D}:=\hat{K}^{-1}\hat{M}
\;\;,
\end{equation}
\begin{equation}
\hat{E}_{\mbox{gh}}^{\mu}:=\hat{K}_{\mbox{gh}}^{-1} \hat{L}_{\mbox{gh}}^{\mu}\;\;,\;\;\hat{D}_{\mbox{gh}}:=\hat{K}_{\mbox{gh}}^{-1}\hat{M}_{\mbox{gh}}\;\;.
\end{equation}
%%%%%%%%%%%%%%%%%%%%%%%%%%%%%%
We use the proper time representation, the expansion with general covariance
and the dimensional regularization \cite{BV}.
The formula of the one loop divergence of the effective action
in the four dimensional case\cite{BOS} is
%%%%%%%%%%%%%%%%%%%%%%%%%%%%%%
\[
\Gamma_{\mbox{div}}^{\mbox{1-loop}}
\]
{\small
\[
= \frac{1}{16 \pi^2 (D-4)} \int d^4 x \sqrt{-g}\left[
\frac{1}{2} \mbox{tr} P^2 
+ \frac{1}{12}\mbox{tr} S_{\lambda \lambda^{\prime}}S^{\lambda \lambda^{\prime}}
-2 \left(
\frac{1}{2} \mbox{tr} P_{\mbox{gh}}^2 
+ \frac{1}{12}\mbox{tr} S_{\mbox{gh};\lambda \lambda^{\prime}}S_{\mbox{gh}}^{\lambda \lambda^{\prime}}
\right)
\right.
\]
}
\begin{equation}
\left.
+ \frac{9 +1 +N -2 \times 4}{180}
\left( R_{\mu \nu \alpha \beta} R^{\mu \nu \alpha \beta}
 - R_{\mu \nu} R^{\mu \nu} \right)
\right]
\end{equation}
%%%%%%%%%%%%%%%%%%%%%%%%%%%%%
where
%%%%%%%%%%%%%%%%%%%%%%%%%%%%%%
\begin{equation}
\hat{P}=\hat{D} + \frac{\hat{1}}{6} R 
-\frac{1}{2}\nabla_{\lambda} \hat{E}^{\lambda}
-\frac{1}{4} \hat{E}^{\lambda}\hat{E}_{\lambda}
\end{equation}
\begin{equation}
\hat{S}_{\lambda \lambda^{\prime}}
=\left[ \nabla_{\lambda}\;,\;\nabla_{\lambda^{\prime}} \right] \hat{1}
+ \frac{1}{2}\left( \nabla_{\lambda} \hat{E}_{\lambda^{\prime}}
                  - \nabla_{\lambda^{\prime}} \hat{E}_{\lambda } \right)
+\frac{1}{4}\left( \hat{E}_{\lambda} \hat{E}_{\lambda^{\prime}}
                  - \hat{E}_{\lambda^{\prime}} \hat{E}_{\lambda } \right)
\end{equation}
$\hat{P}_{\mbox{gh}}$ and $\hat{S}_{\mbox{gh};\lambda \lambda^{\prime}}$
are defined with the same rule.
%\vspace*{\fill}
%\pagebreak
\section{Figure: one loop renormalizable region}
%\vspace*{\fill}
\begin{figure}[htb]
  \parbox{\halftext}{%   %\def\halftext{.471\textwidth}
        \epsfxsize = 8.2cm 
        \centerline{\epsfbox{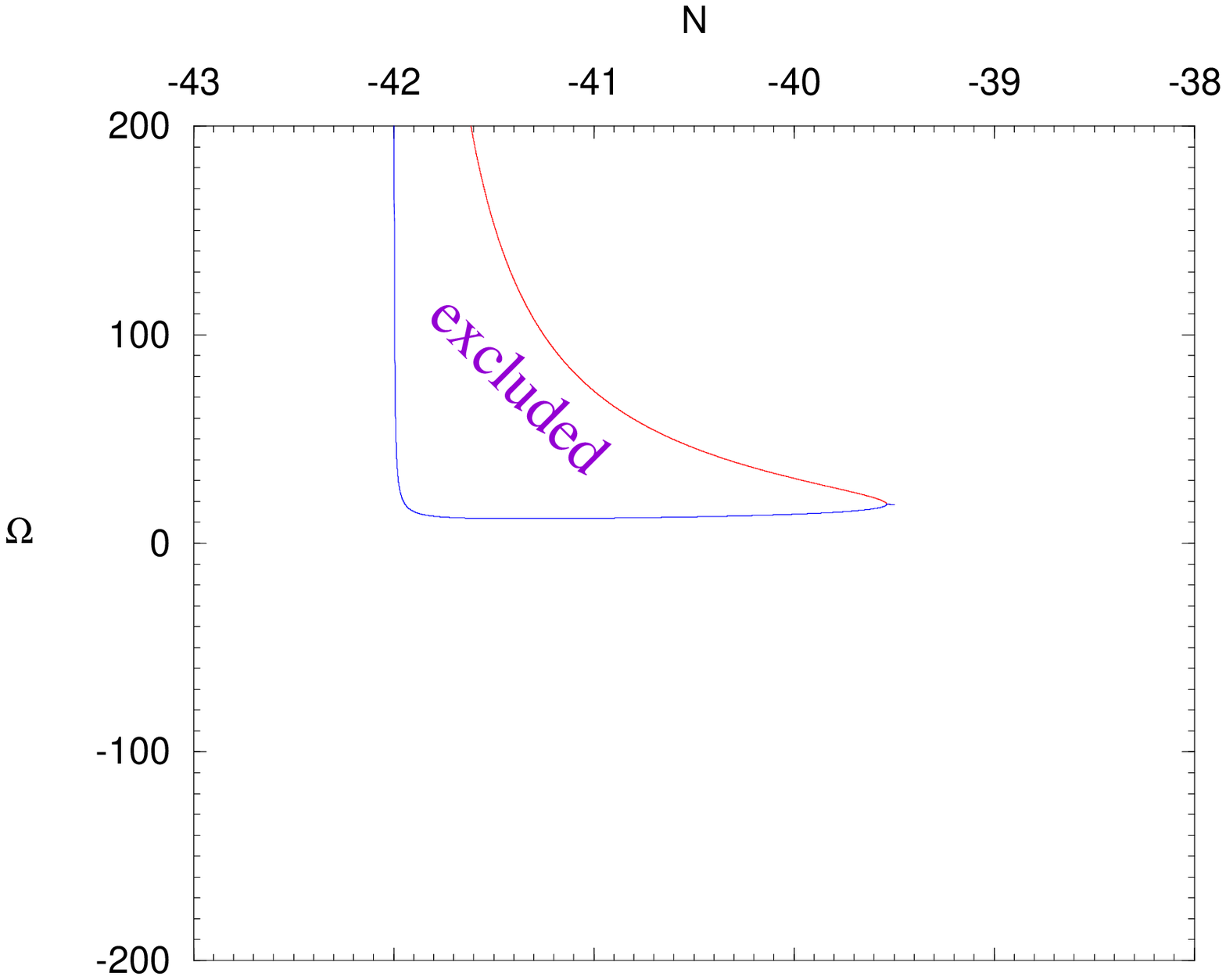}}
        }
  %\hspace{8mm}
  \parbox{\halftext}{
        \epsfxsize = 8.2cm 
        \centerline{\epsfbox{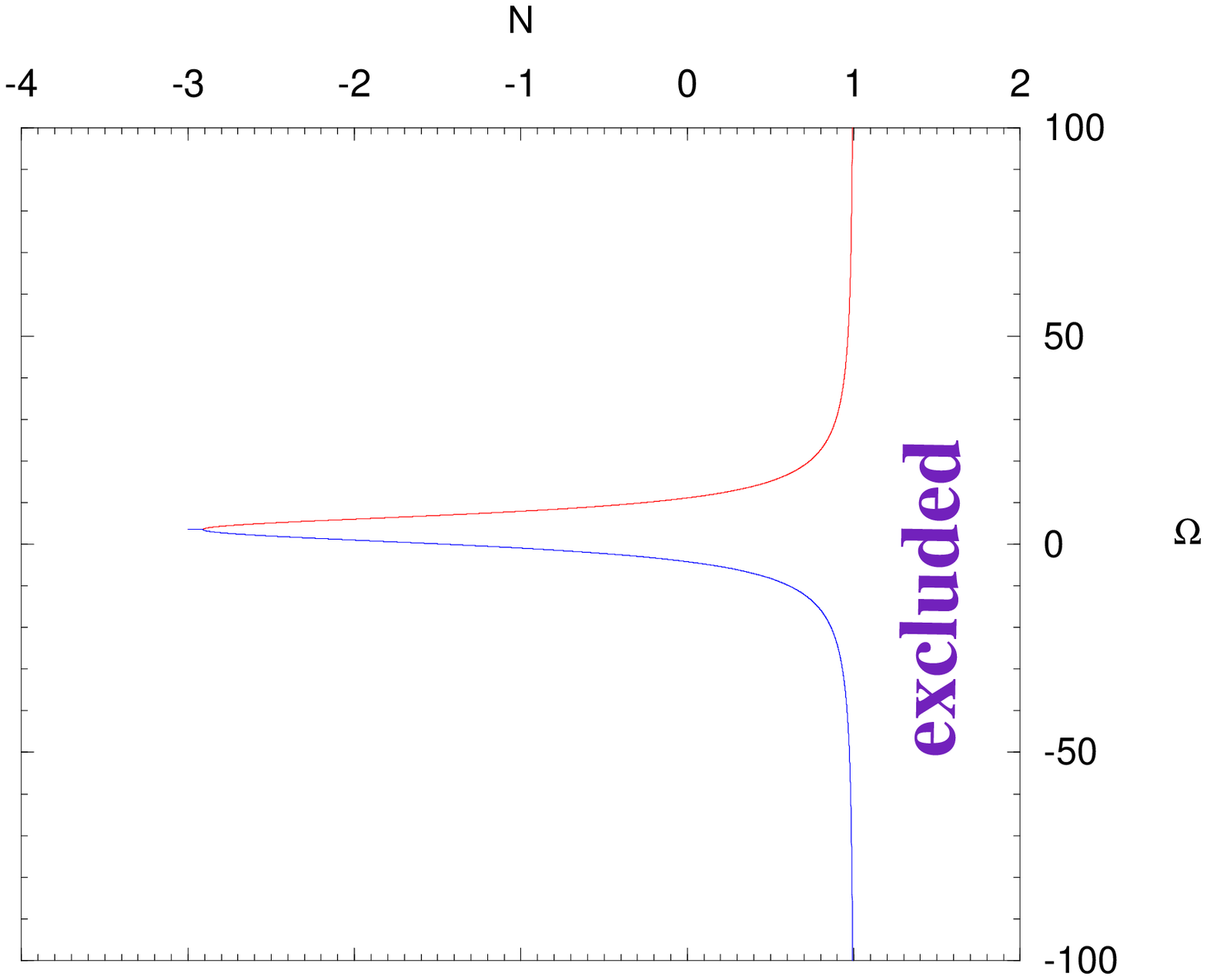}}
          } 
\end{figure}
\vspace*{-2cm}
\begin{figure}[htb]
  \parbox{\halftext}{%   %\def\halftext{.471\textwidth}
        \epsfxsize = 8.2cm 
        \centerline{\epsfbox{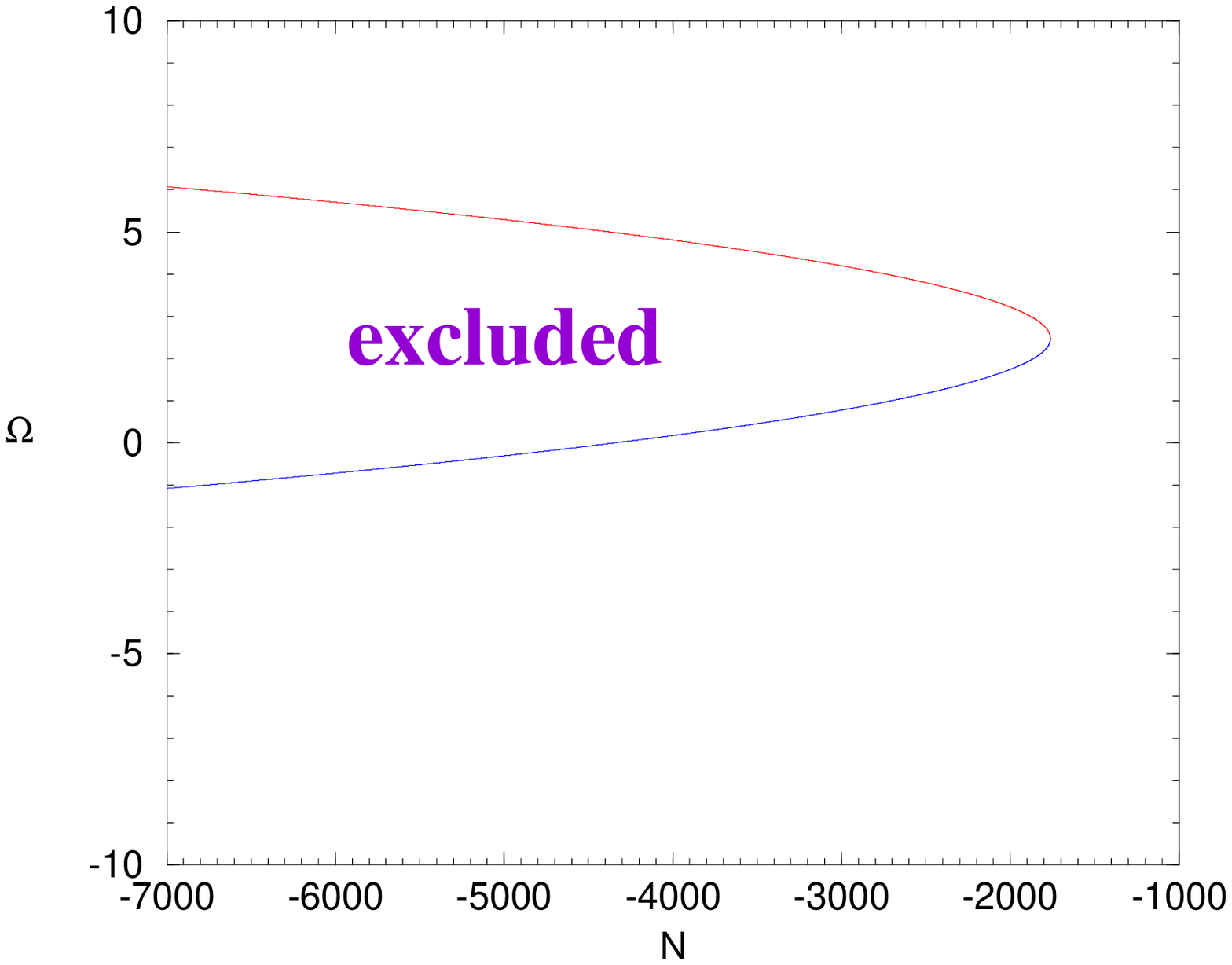}}
          }
  %\hspace{8mm}
  \parbox{\halftext}{
        \epsfxsize = 8.2cm 
        \centerline{\epsfbox{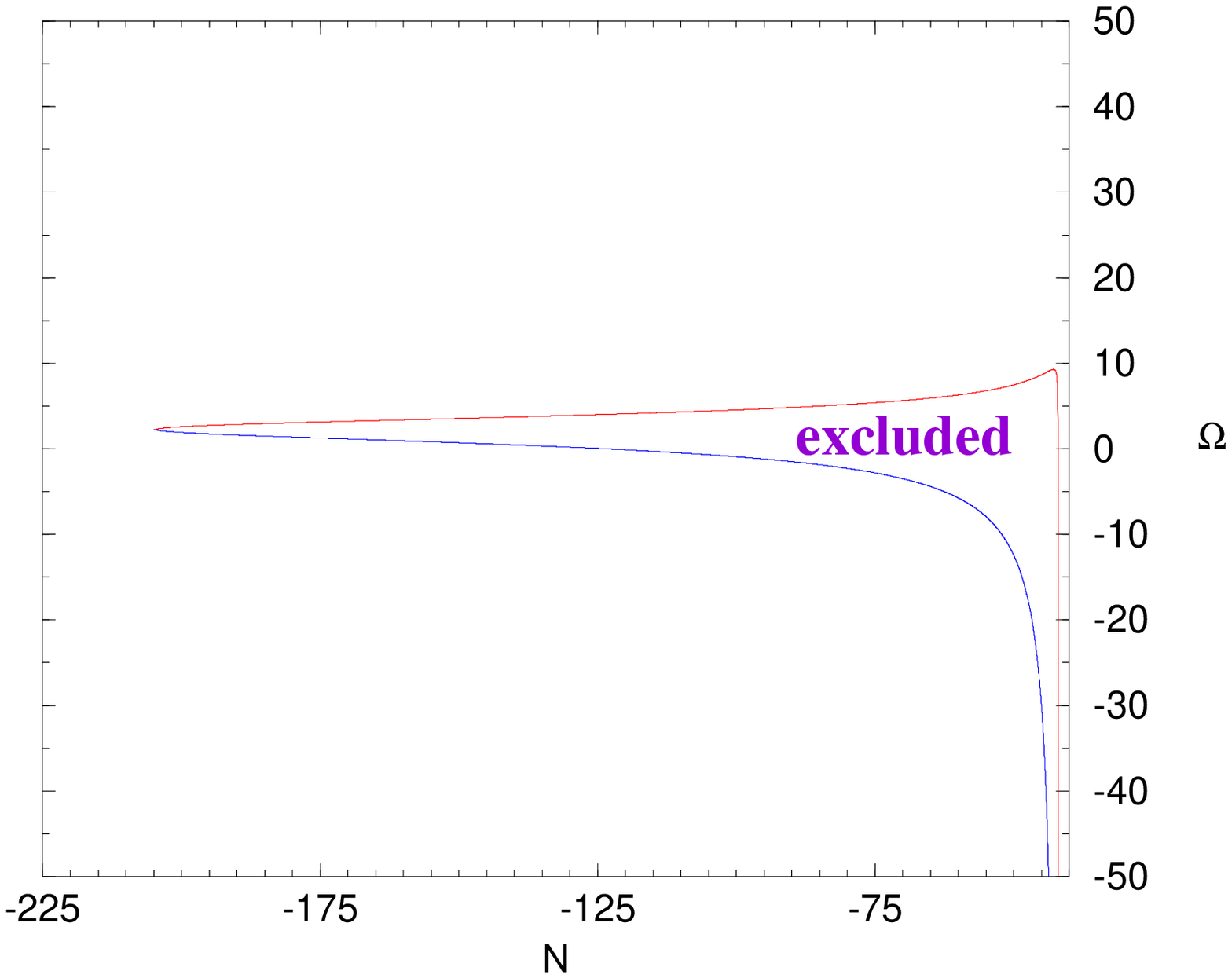}}
          }
      \caption{ 
This  is a region the coefficient of the nonrenormalizable term (4.3) vanishes for real $m$, where $m=m(\phi)$ and $\Omega=\Omega(\phi)$ are defined as
$\frac{B}{\phi B^{\prime}} \equiv \frac{1}{m+2} \Omega $,  
$\frac{B B^{\prime \prime}}{B^{\prime 2}} \equiv \frac{m+1}{m+2}\Omega $.
An interesting example is $B=\mbox{constant}\; \phi^{m+2}$ ($m$ is constant of $\phi$).
This  is the case  in Ref [7].
For an arbitrary value of $\phi$,  $\Omega=1$ and some definite value of $m$, 
this model has no $R^2$ term if $N < 0$.}
\label{fig}
\end{figure}
%%%%%%%%%%%%%%%%%%%%%%%%%%%%%%%%%%%%%%%%%%%%%%%%%%%%%%%%%%%%%%%%%%%%%%%%%%%%%%%%%
%%One loop renormalizable region                                                %
%%in N-dialton and a metric,$A=3/2 B_1^2/B$ case.                               %
%%\[                                                                            %
%%(-121800 + 116800*X+ 4950*X^2 + 50*X^3                                        %
%%\pm ((42 + X)^2                                                               %
%%*(41422416 - 25911984*X - 15066808*X^2 - 441616*X^3 - 2007*X^4 - X^5))^(1/2)) %
%%/(20*(-1 + X)*(42 + X)^2)                                                     %
%%\]                                                                            %
%%%%%%%%%%%%%%%%%%%%%%%%%%%%%%%%%%%%%%%%%%%%%%%%%%%%%%%%%%%%%%%%%%%%%%%%%%%%%%%%%
\pagebreak


\begin{thebibliography}{99}
%%%%%%%%%%%%%%%%%%%%%%%%%%%%%%%%%%%%%%%%%%%%%%%%%%%%%%%%%%%%%
% Some macros are available for the bibliography:
%   o for general use
%      \JL : general journals          \andvol : Vol (Year) Page
%   o for individual journal 
%      \PR  : Phys. Rev.               \PRL : Phys. Rev. Lett.
%      \NP  : Nucl. Phys.              \PL  : Phys. Lett.
%      \JMP : J. Math. Phys.           \CMP : Commun. Math. Phys.
%      \PTP : Prog. Theor. Phys.       \JPSJ: J. Phys. Soc. Jpn.
%      \JP  : J. of Phys.              \NC  : Nouvo Cim.
%      \IJMP: Int. J. Mod. Phys.       \ANN : Ann. of Phys.
% Usage:
%   \PR{D45,1990,345}            ==> Phys.~Rev.\ {\bf D45} (1990), 345
%   \JL{Phys.~Lett.,A30,1981,56} ==> Phys.~Lett.\ {\bf A30} (1981), 56
%   \andvol{B123,1995,1020}      ==> {\bf B123} (1995), 1020
%%%%%%%%%%%%%%%%%%%%%%%%%%%%%%%%%%%%%%%%%%%%%%%%%%%%%%%%%%%%%
\bibitem{GSW} M. B. Green, J. H. Schwarz and E. Witten,
{\sl Superstring Theory},
(Cambridge University Press, Cambridge, 1987).

\bibitem{St} K. S. Stelle,
{\sl Renormalization of the Higher Derivative Quantum Gravity},
Phys. Rev. {\bf 16D}, (1977), 953.

\bibitem{BOS} I. L. Buchbinder, S. D. Odintsov and I. L. Shapiro,
{\sl Effective Action in Quantum Gravity},
(IOP, Bristol, 1992).

\bibitem{HV}G. t'Hooft and M. Veltman,
{\sl One Loop Divergences in the Theory of Gravitation},
Ann. Inst. H. Poincare. {\bf A20}, 69, (1974).

\bibitem{CD} S. Christensen and M. Duff,
{\sl Quantum Gravity with A Cosmological Constant},
Nucl. Phys. {\bf 170B}, (1980), 480.

\bibitem{BKK}  A. O. Barvinski, A. Kamenschik and B. Karmazin,
{\sl The Renormalization Group for Non-Renormalizable Theories: Einstein gravity with A Scalar Field},
Phys. Rev. {\bf D48}, 3677, (1993).

\bibitem{ST1} I. L. Shapiro and H. Takata,
{\sl One Loop Renormalization of the Four-Dimensional Theory for Quantum Dilaton Gravity},
Phys. Rev.{\bf D52}, 2162, (1995).

\bibitem{ST3} I. L. Shapiro and H. Takata,
in Preparation.

\bibitem{Ta1} H. Takata,
{\sl 4D Quantum Dilaton Gravity During Inflation and Renormalization at One-Loop},
hep-th.9604170, KEK-TH-481, KEK-preprint-96-10,  HUPD-9607.

\bibitem{MT} S. Mukaigawa and H. Takata,
hep-th.9605???, KEK-TH-???, KEK-preprint-96-??,  HUPD-9609.

\bibitem{ST2} I. L. Shapiro and H. Takata,
{\sl Conformal Transformation in Gravity},
Phys. Lett. {\bf B361}, (1995), 31.

\bibitem{Ab} L. F. Abbott,
{\sl The Background Field Method Beyond One Loop},
Nucl. Phys. {\bf 185B}, (1981), 189.

\bibitem{De} B. S. DeWitt,
{\sl Dynamical Theory of Groups and Fields},
(Gordon and Breach, NY, 1965).

\bibitem{BV} A. O. Barvinski and G. A. Vilkovisky,
{\sl The Generalized Schwinger-DeWitt Technique in Gauge Theories and Quantum Gravity},
Phys.  Rep. {\bf 119}, No. 1, (1985), 1.

\end{thebibliography}
\end{document}